\documentclass[hyperref]{article}
\usepackage{makeidx}
\usepackage[colorlinks=true,citecolor=black]{hyperref}
\usepackage{hyperref}
\usepackage{multirow}
\usepackage{multicol}
\usepackage[dvipsnames,svgnames,table]{xcolor}
\usepackage{graphicx}
\usepackage{epstopdf}
\usepackage{ulem}
\usepackage{hyperref}
\usepackage{amsmath}
\usepackage{amssymb}
\usepackage{enumerate}
\usepackage[a4paper]{geometry}
\usepackage{subfigure}
\usepackage{caption}
\usepackage{tabulary}
\usepackage{makecell}
\makeatother 
\hypersetup{
colorlinks=true,
linkcolor=black
}
\begin{document}
\captionsetup[figure]{labelfont={bf},name={Fig.},labelsep=period}
\title{\bf\Large Analyzing DNA Hybridization via machine learning}

\date{}
\author{\large Weijun ZHU\\
{\sffamily\small School of Information Engineering, Zhengzhou University, Zhengzhou, 450001 China}
}
\maketitle
{\noindent\small{\bf Abstract:}
  In DNA computing, it is impossible to decide whether a specific hybridization among complex DNA molecules is effective or not within acceptable time. In order to address this common problem, we introduce a new method based on the machine learning technique. First, a sample set is employed to train the Boosted Tree (BT) algorithm, and the corresponding model is obtained. Second, this model is used to predict classification results of molecular hybridizations. The experiments show that the average accuracy of the new method is over 94.2\%, and its average efficiency is over 90839 times higher than that of the existing method. These results indicate that the new method can quickly and accurately determine the biological effectiveness of molecular hybridization for a given DNA design. }

\vspace{1ex}

\section{Introduction}
Real molecular biological experiments are very important in the fields of bioinformatics and DNA (deoxyribonucleic acid) computing. The DNA code design is extremely vital to avoid the uncertainty of the results caused by biological reasons. Hybridization is a molecular biological mechanism often used in DNA algorithms. And its biological effectiveness is often directly related to the success or failure of DNA computing. However, the economic and time cost of real molecular biological experiments is too high, which restricts the application of DNA computing. Therefore, simulated experiments are often used instead of real biological experiments.

However, the time complexity shows an exponential growth with the increase of the length of the molecules, especially the numbers of molecules in the computer simulation of molecular hybridization experiments. Affected by it, the slightly larger simulation of the hybridization is difficult to implement. Unfortunately, this is determined by the inherent complexity of the problem, so that no solution within the framework of hard computing exists. It is well known that computations have the following properties: uncertainties, inaccuracies, incomplete true value, low-cost and robust, in soft computing. Is there any better solution in soft computing? This is the open issue. Motivated by it, we will conduct a study in this paper.

\section{Preliminary}
\subsection{DNA design \& analysis with NUPACK}

DNA encoding, i.e., the optimal design of DNA sequences, is a key to DNA computing. The quality of the codes directly determines the reliability of the DNA computing and affects the efficiency of the biochemical reaction during the hybridization. The success of biochemical experiments depends on the rationality of codes. In addition to meet the requirement of certain DNA computational models and the corresponding DNA algorithms, the designed DNA codes need to meet certain molecular biological constraints, including physical constraints and thermodynamic constraints, such as free energy changes, melting temperature, composition of DNA molecules, coding distance and enzymes etc, in order to ensure specific hybridization. There are many studies on DNA encoding, see \cite{5}, \cite{6}, \cite{7}, \cite{8}, \cite{9}, \cite{10}, \cite{11}, \cite{12}, \cite{13}, \cite{14}, \cite{15} for more details.

The free energy G is the chemical energy released by the hybridization reaction between any two DNA molecules. It releases energy when forming double-stranded molecules and absorbs the same amount of energy when melting. The greater the change in free energy, i.e., $\Delta$G, the more thermodynamically stable the reaction is. Therefore, it will be great DNA coding, if a large amount of energy is released when total double strands occurs after hybridization, and a smaller amount of energy is released when incomplete double strands occurs after hybridization. Furthermore, the melting temperature can also affect the results of hybridization.

NUPACK is developed by Caltech at California Institute of Technology \cite{1}. It is a free tool for analyzing and designing DNA sequence codes. The tool currently enables design of DNA codes meeting a given requirement, and analysis of properties of base-pairing under thermodynamic equilibrium states. The design function of NUPACK can give the optimal molecular coding sequence corresponding to the structure. The analysis function of NUPACK allows the user to simulate the hybridization reaction of DNA molecules at different temperatures and different concentrations, according to the molecular thermodynamic theory. The input of simulation is the reactant molecules and their concentrations under given conditions, and its output is the product molecules and their concentrations under the thermodynamic equilibrium state.
\subsection{ BT algorithm and Graph Lab}
A core goal of Machine Learning (ML) is to classify data. There are many ways applied for classification, such as support vector machines, logistic regression, random forests, decision trees, Boosted Trees (BT) and deep learning. As a class of popular ML algorithms, BT has the following advantages: good effect, insensitive to input and low computational complexity. Thus, BT has been applied to many fields, such as text segmentation \cite{16}, face detection \cite{17}, hand pose recognition \cite{18}, multi-view, multi-pose object detection \cite{19} and emotion recognition \cite{20}, etc. In this paper, we use a BT algorithm called Gradient Boosted Regression Trees (GBRT) \cite{2} to conduct our studies.

As one of the most effective ML algorithms, the GBRT algorithm has the strong generalization ability. GBRT will generate multiple decision trees, and the results of all the trees are accumulated to form the final answer. The core of this algorithm is that each tree learns from the residuals of all previous trees. GBRT can be employed to deal with not only some regression problems but also some binary classification problems. If the latter problem is dealt with, a threshold will be set. A logical 1 will be gotten if the value of regression computation is greater than the threshold. Otherwise, a logical 0 will be gotten.

The advantage of GBRT lies in \cite{3}:
\begin{enumerate}[(1)]
\item Strong ability in handling mixed type of data;
\item Strong predictive ability;
\item Strong robustness against outliers.
\end{enumerate}

The disadvantage of GBRT is that \cite{3} parallel processing cannot be performed. 

Graph Lab is an open source ML package \cite{4}, which was developed by Carnegie Mellon University. This tool integrates a variety of ML algorithms including GBRT, which greatly simplifies the training process of the model, and facilitates users' operations and implementation of specific ML algorithms.

\section{The principle of the new method}
We consider the following specific problem. How to determine whether effective specific hybridization will happen or not, giving a group of complex molecules. The term ``effective'' refers to biologically effective. That is to say, a specific hybridization is effective if and only if the true positive rate (TPR) of the specific hybridization is over 98\%.

\begin{figure}
\centering
\subfigure[compute specific hybridization among one group of molecules, and determine whether the specific hybridization is effective in biology or not, i.e., the ratio of the specific hybridization is over 98 \% ]{
\begin{minipage}[b]{1\textwidth}
\includegraphics[width=\textwidth]{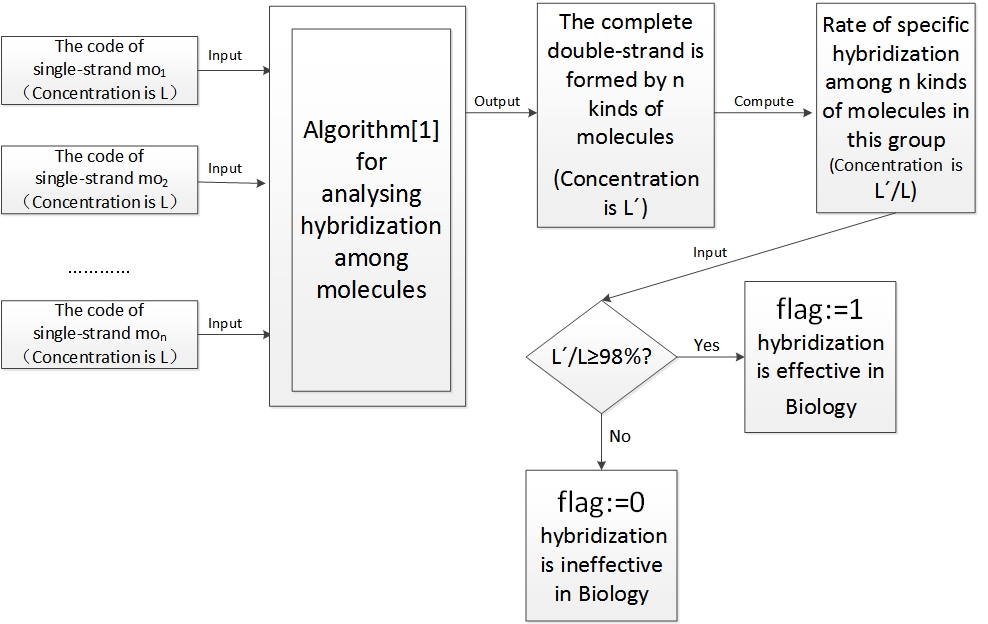} \\
\label{fig1-1}
\end{minipage}
}

\subfigure[ the model M can predict the effectiveness of hybridization for $m_2-m_1$ groups of molecules, since M is obtained by training $m_1$ groups of molecules and their effectiveness of hybridization ]
{
\begin{minipage}[b]{1\textwidth}
\includegraphics[width=\textwidth]{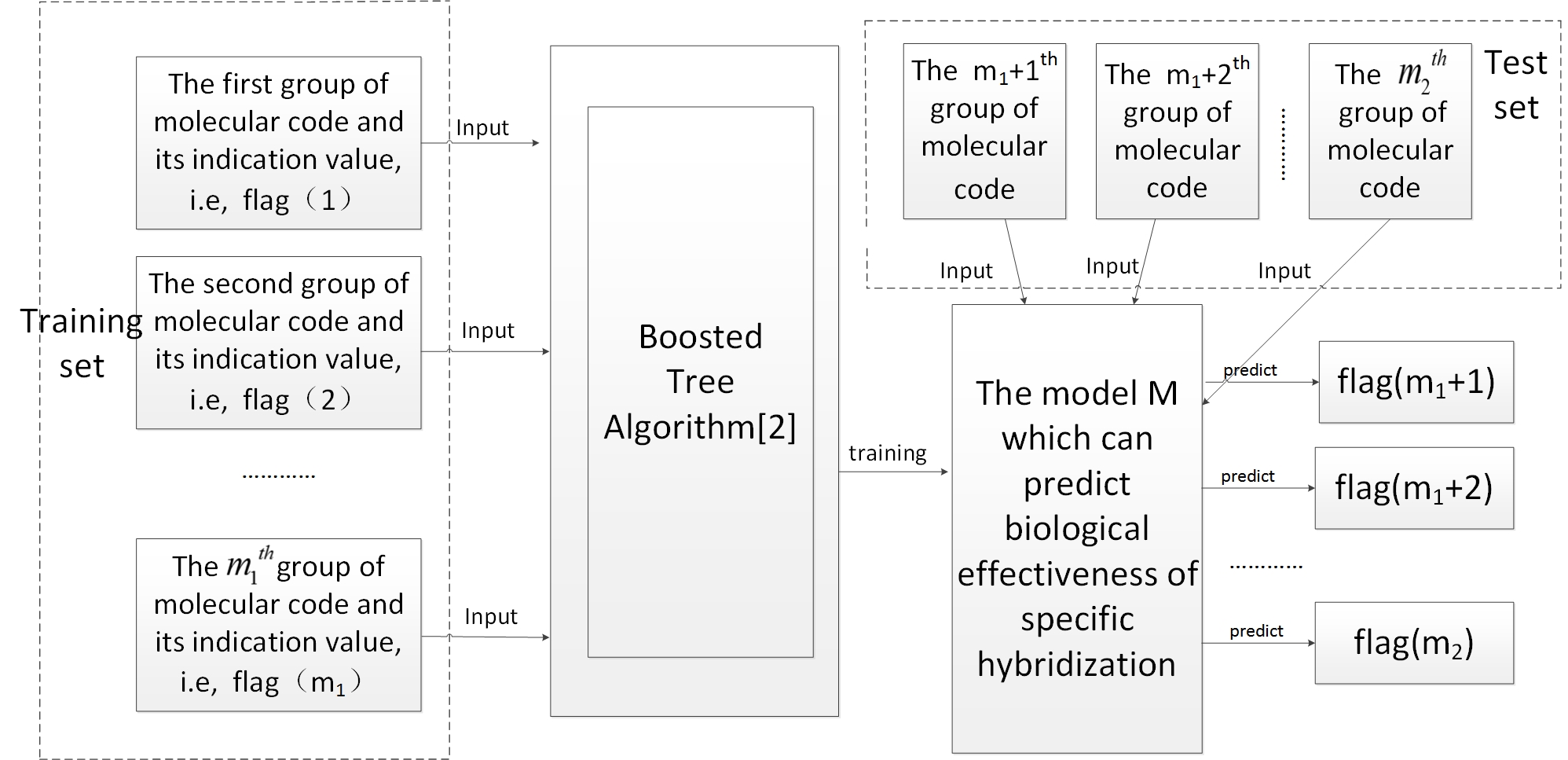} \\
\label{fig1-2}
\end{minipage}
}

\caption{given one group of molecular codes, the new method can determine whether the specific hybridization among these molecules is effective in biology or not }
\label{fig1}
\end{figure}

The principle of our method is shown in Fig. \ref{fig1}. The core is to train a large number of records containing information on the codes of DNA molecules and their results of specific hybridization, using BT algorithm. Thus, a ML model called M which has a predictive ability is obtained. And then, a given set of DNA molecular codes is input into the model M. As a result, the output of M is the predicted result in terms of the effectiveness of specific hybridization among the molecules.

The steps of the process can be described as follows.
\begin{enumerate}[(1)]
\item As shown in Fig. \ref{fig1-1}, one can analyze and get the double-stranded DNA molecules and their concentrations which are produced by specific hybridization among n kinds of reactant molecules with a suitable initial concentration after it anneals to a target temperature. On the basis of it, one can perform a binary classification for the biologic effectiveness of specific hybridization, according to the value of concentrations of target molecules.
\item Step (1) is repeated $m_1$ times, and a training set containing $m_1$ records is gotten. See the left part of Fig. \ref{fig1-1}.
\item Train the BT algorithm using the training set obtained in step (2), and the model M is obtained. See the middle part of Fig. \ref{fig1-2}.
\item Input a set of DNA molecular codes into the trained model M. Whether the reactant molecules can effectively hybridize specifically to obtain the target molecules? It can be predicted.
\end{enumerate}
\section{Simulated experiments}
\subsection{The objective of the experiments}
We will explore the ability and the efficiency of the new method based on ML and the analytic algorithm in NUPACK, in terms of the biologic effectiveness of specific hybridization. Specifically, can the former method improve the analytical efficiency significantly under the premise that the former method can approach the latter one in terms of analytic ability?
\subsection{The simulation platform}
\begin{enumerate}[(1)]
\item CPU: Intel(R) Core(TM) i7-4770 CPU @ 3.40GHz.
\item RAM: 16.0G RAM.
\item OS: Windows 10.
\item NUPACK: for designing DNA molecules and analyzing hybridizations.
\item Graph Lab: for implementing the BT algorithm.
\end{enumerate}

\subsection{ Experimental Procedures}
\begin{enumerate}[(1)]
\item $m_2/2$ groups of DNA molecules are designed using NUPACK. Each group consists of following four kinds of molecules: one kind of long single-stranded molecule with 36 bases and three kinds of short single-stranded molecules with 8, 20, and 8 bases respectively. The one kind of long single-stranded molecule and the three kinds of short single-stranded molecules form a complete WC complement at the logic level.
\item  NUPACK is applied to analyze hybridization for each one of $m_2/2$ groups of DNA molecules mentioned in (1) one by one. The initial concentration is set to 100uM, and the temperature is reduced to 23 degrees Celsius by anneal. Hybridization is performed to obtain the concentration of the complete double-stranded molecules, and the NUPACK analytic time is also recorded.
\item  For each group of molecules, a record consisting of five fields is obtained according to the four kinds of molecular codes mentioned in (1) and the concentration of the complete double-stranded molecules mentioned in (2). In this record, the first four fields note the four kinds of DNA molecular codes, and the last field notes ``1", if the ratio of the concentration of the complete double-stranded molecules to 100uM is greater than or equal to 98\%, otherwise, the last field notes ``0". 
\item According to the relevant information on the $m_2/2$ groups of molecules mentioned in (1), one can get the corresponding $m_2/2$ groups of records. On the basis of it, the average NUPACK analytic time for all these $m_2/2$ groups can be computed.
\item $m_2/2$ groups of DNA molecules are designed using a random algorithm/tool. Each group consists of following four kinds of molecules: one kind of long single-stranded molecule with 36 bases and three kinds of short single-stranded molecules with 8, 20, and 8 bases respectively. The one kind of long single-stranded molecule and the three kinds of short single-stranded molecules form a complete WC complement at the logic level.
\item NUPACK is applied to analyze hybridization for each one of $m_2/2$ groups of DNA molecules mentioned in (5) one by one. The initial concentration is set to 100uM, and the temperature is reduced to 23 degrees Celsius by anneal. Hybridization is performed to obtain the concentration of the complete double-stranded molecules, and the NUPACK analytic time is also recorded.
\item For each group of molecules, a record consisting of five fields is obtained according to the four kinds of molecular codes mentioned in (5) and the concentration of the complete double-stranded molecules mentioned in (6). In this record, the first four fields note the four kinds of DNA molecular codes, and the last field notes ``1", if the ratio of the concentration of the complete double-stranded molecules to 100uM is greater than or equal to 98\%, otherwise, the last field notes ``0".
\item According to the relevant information on the $m_2/2$ groups of molecules mentioned in (5), one can get the corresponding $m_2/2$ groups of records. On the basis of it, the average NUPACK analytic time for all these $m_2/2$ groups can be computed.
\item According to the average analytic time mentioned in (4) and the average analytic time mentioned in (8), the NUPACK average analytic time for all $m_2$ groups of molecular hybridizations can be computed.
\item All $m_2$ records obtained in (4) and (8) are used as the training set and test set of BT algorithm on Graph Lab, where the training set has $m_1$ records. Thus, this is a binary classification problem. The model M will predict the value of the fifth field, i.e., ``1" or ``0", since the first four fields have been given. 
\item The result of prediction and the consumed time for each record in the test set are noted. On the basis of it, the average predictive accuracy and the average time for all test records can be computed.
\item One can compare the BT ability and the NUPACK ability in terms of analysis of specific hybridizations, according to the average predictive accuracy mentioned in (11).
\item With the average analytic time of NUPACK mentioned in (9) and the average predictive time of BT mentioned in (12) at hands, one can compare the efficiency of the BT algorithm to the one of NUPACK.  
\end{enumerate}
\subsection{Experimental results }
\subsubsection{NUPACK experiments}
In section 4.4.1, we have $n=4, m_2=810$. There are 406 groups of DNA molecules designed by the manually operated random algorithms, and 404 groups of DNA molecules designed by NUPACK. All these 810 groups of DNA molecules will be analyzed by NUPACK for their specific hybridization. In section 4.4, the specific hybridization is defined as follows: a kind of long single strand in a group of DNA molecules and all kinds of short single strands in this group of DNA molecules are paired to form a kind of complete double strand with the Watson-Crick complementary way.

\begin{table}
\centering
\begin{footnotesize}
		\begin{tabular}{|c|c|c|c|c|c|c|c|c|c|} \hline
			Group&Strand1&Strand2&Strand3&Strand4&\makecell{Analysis \\ time}&\makecell{Concentration \\ after specific\\ hybridization}&TPR&\makecell{Result of \\classification}&\makecell{Design\\ manually\\ or by\\ NUPACK?}\\\hline
			\makecell{1}& \makecell{CGCG\\GCGC\\CGCA\\GCGG\\TGCA\\CACG\\CGCG\\CGGC\\GGCG} & \makecell{CGCC\\GCCG}& \makecell{CGCG\\CGTG\\TGCA\\CCGC\\TGCG}&\makecell{GCGC\\CGCG}  &12s&99uM&\makecell{99\%\\(=99\\uM\\/100\\uM)}&\makecell{1\\(since 99\% \\$\ge$98\% )}&NUPACK\\\hline
			2& \makecell{CTAC\\GTAG\\CCTC\\CATA\\GAGA\\CTTG\\GAAT\\AGAC\\GTCT} & \makecell{AGAC\\GTCT}& \makecell{ATTC\\CAAG\\TCTC\\TATG\\GAGG}&\makecell{CTAC\\GTAG}  &15s&68uM&\makecell{68\%\\(=68\\uM\\/100\\uM)}&\makecell{0 \\(since 68\%\\$<$98\% )}&manually\\\hline
		\end{tabular}
		\caption{part of experiments via NUPACK: analyzing two groups of molecules as an example}
		\label{tab1}
\end{footnotesize}
	\end{table}

Table \ref{tab1} shows the two groups of instances. Group 1 is one of the 404 groups of DNA molecules which are designed via NUPACK, and Group 2 is one of the 406 groups of DNA molecules which are designed artificially. The second field, the third one, the fourth one and the fifth one in Table \ref{tab1} show the first kind of molecular codes, the second one, the third one and the fourth one for each group of DNA molecules, respectively.

\begin{figure}[!htb]
\centering
\subfigure[base pairing among molecules: positions and probabilities ]{
\begin{minipage}[b]{0.45\textwidth}
\includegraphics[width=\textwidth]{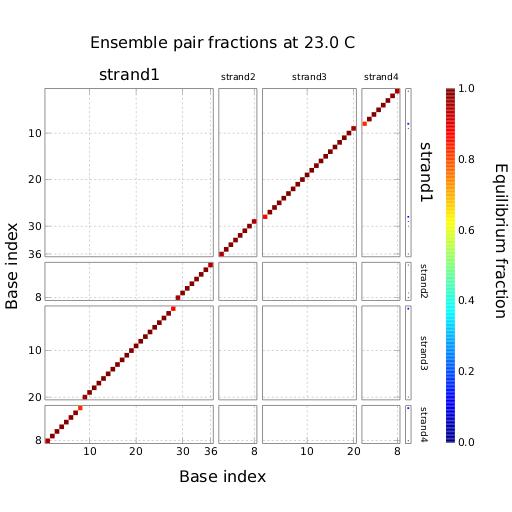} \\
\label{fig2-1}
\end{minipage}
}
\subfigure[ the different molecular concentrations of different products]
{
\begin{minipage}[b]{0.45\textwidth}
\includegraphics[width=\textwidth]{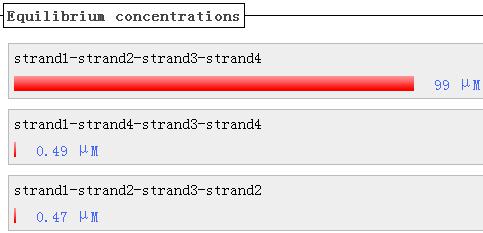} \\
\label{fig2-2}
\end{minipage}
}

\caption{the result of hybridization for Group 1 }
\label{fig2}
\end{figure}

Fig. \ref{fig2} illustrates the results of the analysis for Group 1 via NUPACK. And Fig. \ref{fig2-1} shows the positions of the base pairing among all 4 kinds of molecules and their pairing probability after hybridization is completed. We find out that the first eight bases in strand 1 are paired with all eight bases in strand 4 in accordance with the order, observing the position of chromatic oblique line in this graph. And the twenty bases in the middle in strand 1 are paired with all twenty bases in strand 3 in accordance with the order, whereas the last eight bases in strand 1 are paired with all eight bases in strand 2 in accordance with the order. As a result, the target double strands are completely formed. Observing the color of the chromatic oblique line in this graph, and comparing it with the change of colors on the right vertical bar in this graph, we find out that all the probabilities of base pairing are almost one hundred percent. This facts indicates that specific hybridization occurs with very high probability.

The kind of molecule called strand 1- strand 2- strand 3- strand 4 in Fig. \ref{fig2-2} is a kind of target molecules originated from the specific hybridization. And the concentration of this kind of target molecules is 99uM, as shown in Fig. \ref{fig2-2}. The probability of specific hybridization, i.e., True Positive Rate (TPR), is 99uM/100uM=99\%, since the initial concentration of each kind of molecules is 100uM. It is obvious that the value of the above TPR is greater than 98\%. As a result, this value is classified as 1, indicating that the specific hybridization in Group 1 is biologically effective.

\begin{figure}[!htb]
\centering
\subfigure[base pairing among molecules: positions and probabilities ]{
\begin{minipage}[b]{0.45\textwidth}
\includegraphics[width=\textwidth]{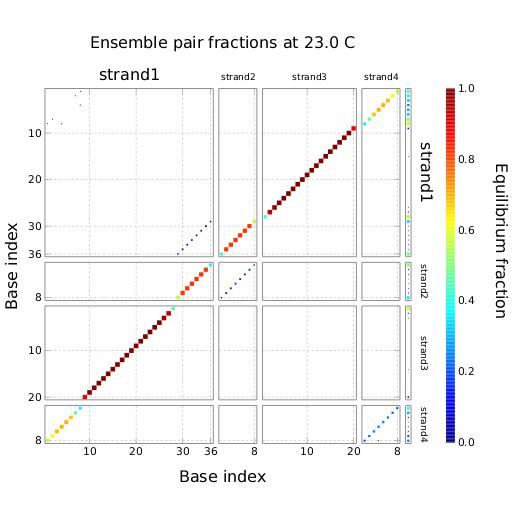} \\
\label{fig3-1}
\end{minipage}
}
\subfigure[ the different molecular concentrations of different products]
{
\begin{minipage}[b]{0.45\textwidth}
\includegraphics[width=\textwidth]{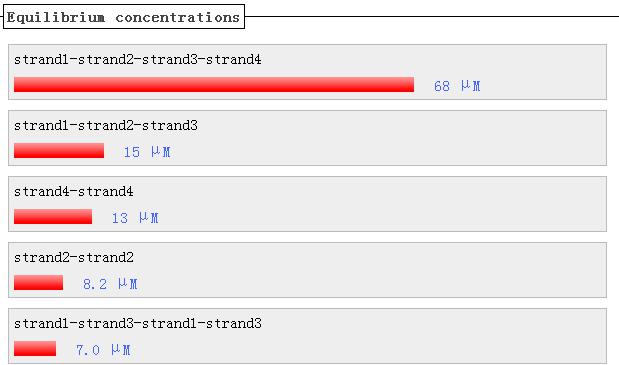} \\
\label{fig3-2}
\end{minipage}
}

\caption{the result of hybridization for Group 2 }
\label{fig3}
\end{figure}

Fig. \ref{fig3} illustrates the results of the analysis for Group 2 via NUPACK. We find out that there are several slants in all bases in strand 2, observing the position of chromatic oblique line in this graph. This phenomenon indicates that the pairing position is difficult to determine. And strand 4 suffers a similar problem. Therefore, the specific hybridization is not formed with high probability. Furthermore, the molecular concentration of strand 1- strand 2- strand 3- strand 4 is only 68uM, according to in Fig. \ref{fig3-2}. The specific hybridization in Group 2 is biologically ineffective since 68uM/100uM=68\% and 68\%$<$98\%. In other words, the probability of specific hybridization is classified as 0.

The sixth fields in Table \ref{tab1} shows the time required for analyzing a group of molecules via NUPACK, whereas the seventh fields indicate the molecular concentration of target products after specific hybridization is completed. And the true positive rate of specific hybridization is shown in the eighth field. Furthermore, the ninth field shows the results of classification based on the true positive rates, whereas the tenth field points out the source of a group, i.e., is it an artificial design or a design via NUPACK?
\subsubsection{Experiments via Graph Lab}

In section 4.4.1, we get $m_2=810$ groups of molecular experimental results via NUPACK analysis. These results are noted as 810 records in a database, which has the same format with Table \ref{tab1}. And these 810 records are original data of the experiments in section 4.4.2. On the one hand, $m_1$ records of all 810 ones are made up a training set, where each record of the training set has and only has the second field, the third one, the fourth one, the fifth one and the ninth one in Table \ref{tab1}. And BT algorithm use this training set to get a model M. On the other hand, $m_2-m_1$ records of all 810 ones are made up a test set, where each record of the test set has and only has the second field, the third one, the fourth one and the fifth one in Table \ref{tab1}. For a test record, the values of these four fields are inputted to M. We can determine whether the result of classification is right or not, by comparing the result of classification, i.e., the output of M, with the value of the ninth field of the record.    

\begin{figure}
\centering
\subfigure[Result of prediction 1: using 205 records in sample set, containing 21 test records]{
\begin{minipage}[b]{0.48\textwidth}
\includegraphics[width=\textwidth]{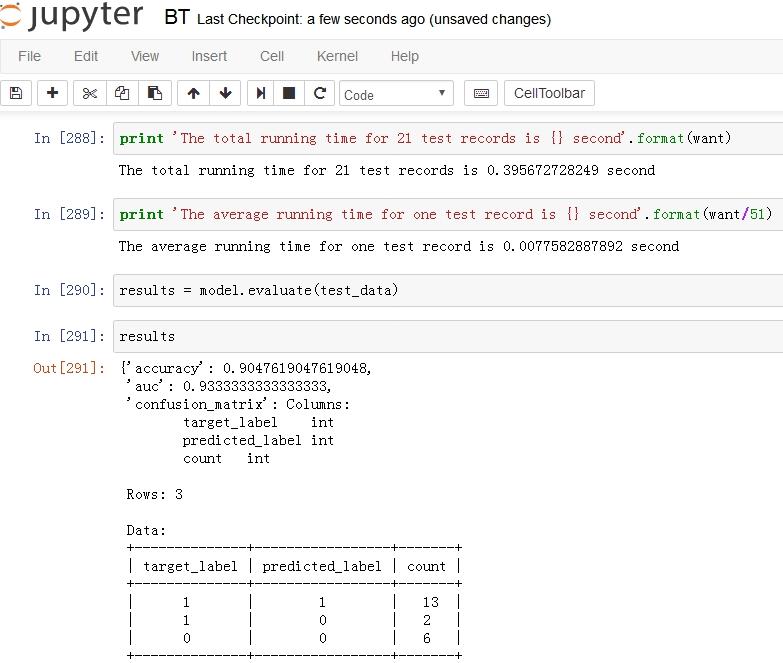} \\
\label{fig4-1}
\end{minipage}
}
\subfigure[Result of prediction 2: using 305 records in sample set, containing 33 test records]{
\begin{minipage}[b]{0.48\textwidth}
\includegraphics[width=\textwidth]{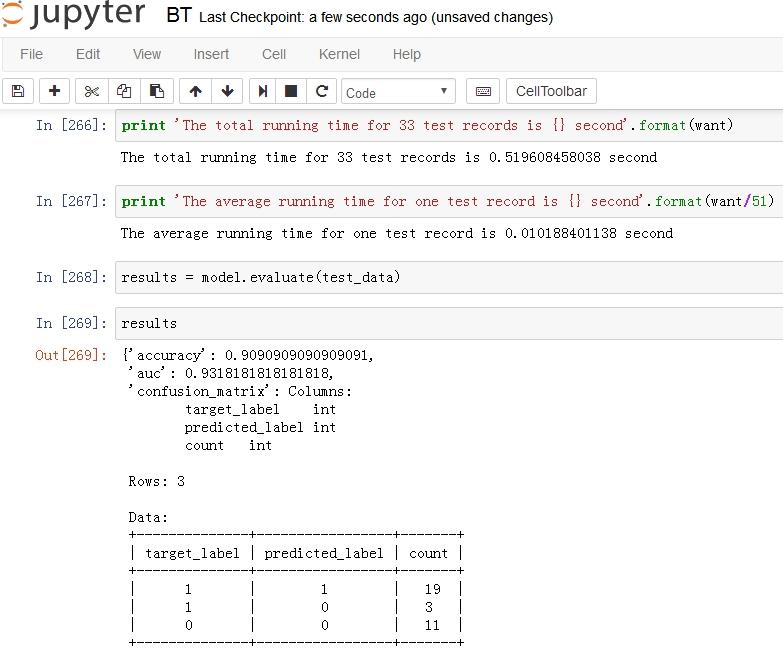} \\
\label{fig4-2}
\end{minipage}
}
\subfigure[Result of prediction 3: using 415 records in sample set, containing 51 test records]{
\begin{minipage}[b]{0.48\textwidth}
\includegraphics[width=\textwidth]{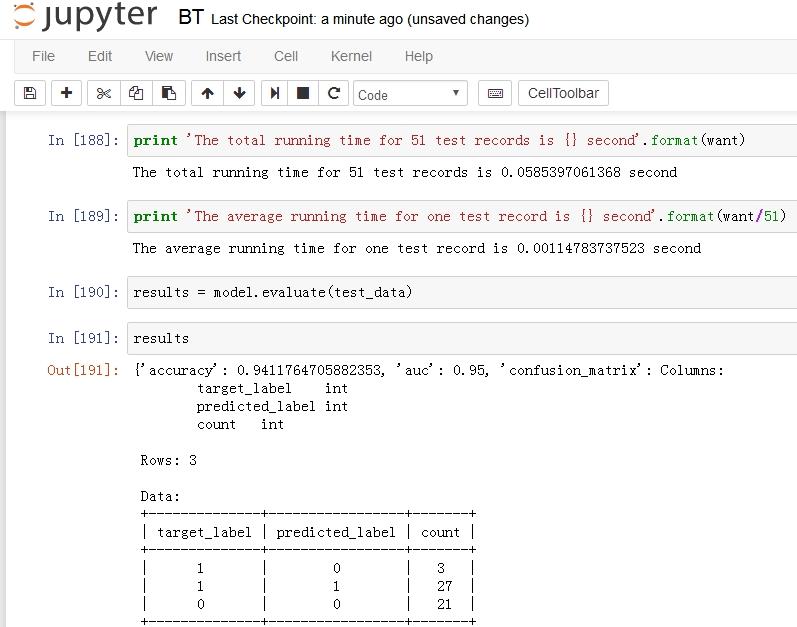} \\
\label{fig4-3}
\end{minipage}
}
\subfigure[Result of prediction 4: using 810 records in sample set, containing 52 test records]{
\begin{minipage}[b]{0.48\textwidth}
\includegraphics[width=\textwidth]{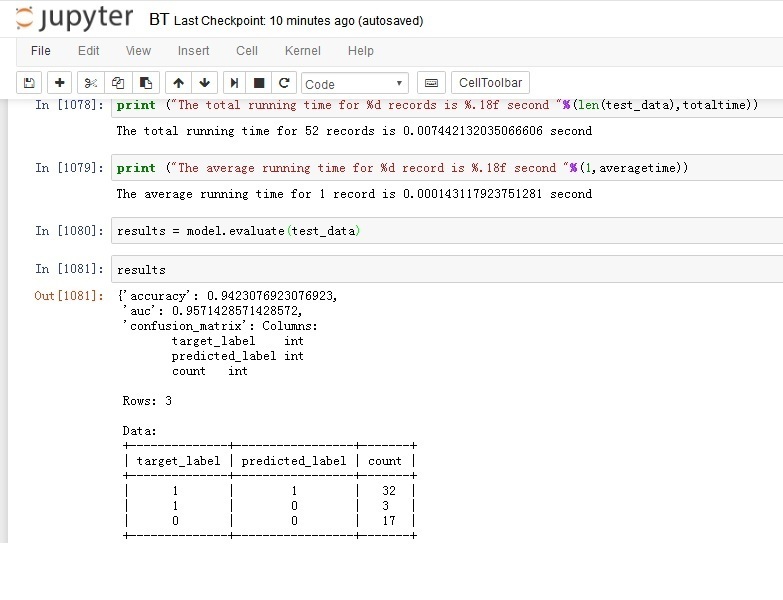} \\
\label{fig4-4}
\end{minipage}
}

\caption{Result of the three groups of prediction based on different numbers of records }
\label{fig4}
\end{figure}

Let the numbers of sample sets, i.e., $m_2$, be 205, 305, 415 and 810, respectively, and the numbers of test sets, i.e., $m_2-m_1$, be 21, 33, 51 and 52, respectively. Four groups of experiments via Graph Lab have been designed. The result of the program is shown in Fig. \ref{fig4}. This figure indicates that the accuracies of prediction are 90.5\%, 90.9\%, 94.1\% and 94.2\%, respectively, as well as the average time required for predicting single record are 0.0078 seconds, 0.0102 seconds, 0.0011 seconds and 0.000143 seconds, respectively, in the above four groups of experiments. Fig. \ref{fig5} illustrates the bar graph indicating these results.

\subsection{Discussion}
\begin{figure}
\centering
\subfigure[accuracy of prediction]{
\begin{minipage}[b]{0.4\textwidth}
\includegraphics[width=\textwidth]{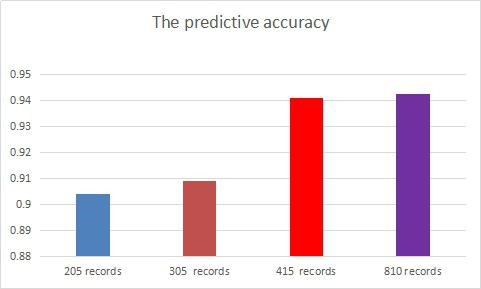} \\
\label{fig5-1}
\end{minipage}
}
\subfigure[average time for prediction with one record]{
\begin{minipage}[b]{0.4\textwidth}
\includegraphics[width=\textwidth]{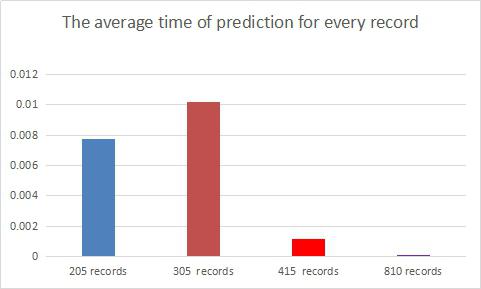} \\
\label{fig5-2}
\end{minipage}
}

\caption{predictions with different numbers of sample records using BT algorithm }
\label{fig5}
\end{figure}

\begin{enumerate}[(1)]
\item First, comparing Fig. \ref{fig2} with Fig. \ref{fig3}, we find out that the true positive rates of specific hybridizations of molecular codes which are designed by NUPACK are significantly higher than that of artificial design molecules. Such experimental results are very common in our 404 groups of molecular codes designed by NUPACK and 406 groups of molecular codes designed manually. There are only 13 groups of molecular codes designed by NUPACK whose TPR is less than 98\%, and these groups made up 3.2\% of total 404 groups designed by NUPACK. In contrast, there are 328 groups of artificial molecular codes whose TPR is less than 98\%, and these groups made up 81\% of total 406 groups designed manually. This phenomenon suggests that DNA molecular codes designed by NUPACK are superior to artificial ones. Meanwhile, this phenomenon also indicates that the sample data from NUPACK provide a large number of ``1" for the BT algorithm, whereas the artificial sample data provide a multitude of ``0" for the BT algorithm. This ensures that the machine can learn the adequate information on positive cases and negative ones, ensuring the generalization ability of the model M.
\item Second, the average time of one group of molecular hybridization analyzed by NUPACK and the one predicted by Graph Lab are presented in Table \ref{tab2}, if the number of sample records is 205, 305, 415 and 810. The table shows that efficiency of the new method based on the BT algorithm increases 90839 times at most, compared with the traditional methods for analyzing DNA molecular hybridizations. The reason for this phenomenon is that the ML-based method does not need to traverse the combinatorial state space, compared with the traditional methods. In fact, the average time of one group of molecular hybridization analyzed by NUPACK increase to 2 minutes if the kinds of single-strand molecules increase to five, i.e., n=5, whereas the average time of one group of molecular hybridization analyzed by NUPACK is over 24 hours if the kinds of single-strand molecules increase to seven, i.e., n=7. Thus, NUPACK provides no service when n=7. The reason for this phenomenon is that the traditional methods for analyzing DNA hybridizations face a problem of exponentially computational complexity. This problem does not exist using the new method. The average predictive time does not change at the order of magnitude, even if there are seven kinds of molecules in one group of DNA molecules which has a total of 50 bases.
\item Three, with the new method at hand, the average accuracy for predicting the analytic results of NUPACK is over 94.2\%, as shown in Fig. \ref{fig5-1}. This phenomenon suggests that the cost of using the new method is acceptable, compared to the significant benefits of using the new method, since the advantage in using the new method is that the analyzing efficiency is improved more than 90839 times, whereas the disadvantage in using the new method is that the analyzing accuracy will go down by only less than 6 percent.
\end{enumerate} 

\begin{table}
\centering
		\begin{tabular}{|c|c|c|c|c|} \hline
		\makecell[tl]{Number of \\ records in \\ sample set} &	\makecell[tl]{Average time required for analyzing \\ hybridization among one group of \\ molecules via NUPACK, i.e., $t_1$ } & \makecell[tl]{Average time required for \\ predicting the result of \\ NUPACK ' s analysis, i.e., $t_2$} &
$t_2 / t_1$  & $t_1 / t_2$ \\ \hline
205 & 12.0097 seconds & 0.0078 seconds & 0.06\%	& 1540 \\ \hline
305&	13.0655 seconds&	0.0102 seconds&	0.08\%	&1281 \\ \hline
415	&13.2867 seconds&	0.0011 seconds&	0.008\%&	12079 \\ \hline
810	&12.99 seconds&	0.000143 seconds&	0.001\%&	90839 \\ \hline
		\end{tabular}
		\caption{The new method improves the efficiency}
		\label{tab2}
	\end{table}

\section{Conclusions}
In this study, a new approach based on BT algorithm is introduced to analyze DNA molecular specific hybridization. To the best of our knowledge, this is the first method for analyzing molecular specific hybridization using the machine learning technique. This is the main contribution of this paper.

The core of the existing methods is to traverse combinatorial state space. And the new method based on BT complements the existing ones. The new method has an acceptable analytical accuracy which has declined slightly in exchange for a substantial increase in analytical efficiency. As a result, the more complex the DNA molecules involved in the hybridization become, the more obvious the comparative advantage of the new methods is. This is the benefit of using the new method.
\section*{Acknowledgements}
This work has been supported by the National Natural Science Foundation of China (No.U1204608, No. 61572444), the National Key Research and Development Program of China (No.2016YFB0800101).

\end{document}